\documentclass[12pt]{article}
\usepackage{amsmath}
\def\a{\alpha}
\def\b{\beta}

\def\g{\gamma}
\def\p{\psi}

\def\t{\theta}

\def\s{\sigma}

\def\be{\begin{equation}}
\def\ee{\end{equation}}
\def\arr{\begin{array}{rll}}
\def\ea{\end{array}}
\def\bea{\begin{eqnarray}}
\def\eea{\end{eqnarray}}

\def\N2{$N{=}2$}

\def\>{\rangle}
\def\<{\langle}
\def\+{\dagger}
\def\={\ =\ }

\begin{document}
\renewcommand{\thefootnote}{\fnsymbol{footnote}}
\begin{titlepage}
\setcounter{page}{0}
\begin{flushright}
LMP-TPU--3/11  \\
\end{flushright}
\vskip 1cm
\begin{center}
{\LARGE\bf  $\mathcal{N}=2$ superparticle near horizon of extreme}\\
\vskip 0.5cm
{\LARGE\bf  Kerr--Newman-AdS-dS black hole}\\
\vskip 2cm
$
\textrm{\Large Anton Galajinsky and Kirill Orekhov\ }
$
\vskip 0.7cm
{\it
Laboratory of Mathematical Physics, Tomsk Polytechnic University, \\
634050 Tomsk, Lenin Ave. 30, Russian Federation} \\
{Emails: galajin@mph.phtd.tpu.ru, orekhov@mph.phtd.tpu.ru}

\end{center}
\vskip 1cm
\begin{abstract} \noindent
Conformal mechanics related to the near horizon extreme Kerr--Newman-AdS-dS black hole
is studied. A unique $\mathcal{N}=2$ supersymmetric extension of the conformal mechanics is constructed.
\end{abstract}

\vspace{0.5cm}

\end{titlepage}

\renewcommand{\thefootnote}{\arabic{footnote}}
\setcounter{footnote}0

\noindent
{\bf 1. Introduction}\\

A large class of extreme black holes in four and five dimensions
exhibits $SO(2,1)$ symmetry in the near horizon limit \cite{klr,cl}.
The most interesting example of such a kind is the Kerr black hole in four dimensions \cite{bh}.
The idea in \cite{str} to consider excitations around the near horizon extreme Kerr geometry and to
extend the corresponding $SO(2,1) \times U(1)$ symmetry group to the Virasoro group of asymptotic
symmetries has paved the way for extensive investigation of the Kerr/CFT correspondence\footnote{By now there is an extensive literature on the subject. For a more complete list of references see e.g. a recent work \cite{aot}.}.

A useful means of studying geometry of vacuum solutions of the Einstein equations is provided by
particle mechanics on a curved background. Apart from the issue of geodesic completeness, the knowledge of conserved charges of a particle helps uncover symmetries of the background. For example, the discovery of a quadratic first integral for
a massive particle in the Kerr space--time \cite{car} preceded the construction of the second rank Killing tensor for the Kerr geometry \cite{wp}.
Because $SO(2,1)$ is the conformal group in one dimension, particle models derived from the near horizon geometries with the $SO(2,1)$ isometry
will automatically be conformal invariant.

Conformal mechanics associated with the near horizon extreme Kerr black hole in four dimensions has been studied recently in \cite{ag}.
It was shown that in the near horizon limit the Killing tensor degenerates into a quadratic combination of Killing vectors corresponding to
the $SO(2,1) \times U(1)$ isometry group. Furthermore, because $SO(2,1) \times U(1)$ is the bosonic subgroup of $SU(1,1|1)$,
an $\mathcal{N}=2$ superconformal extension is feasible. While a relation between first integrals of a particle propagating on a
curved background and Killing vectors characterizing a background geometry is well known, a link between supersymmetry charges and Killing spinors is poorly understood. A possibility to construct
$\mathcal{N}=2$ superparticle moving near the horizon of the extreme Kerr black hole thus raises the question about
geometric interpretation of the supercharges.

The purpose of this work is to extend the analysis in \cite{ag} to the case of the near horizon extreme Kerr--Newman-AdS-dS black hole \cite{strominger}. There are two motivations for this study. First, the analysis in \cite{kp,ck} indicates that the extreme Kerr--Newman-AdS black hole is supersymmetric. Then one can hope to relate supersymmetry generators of an $\mathcal{N}=2$ superparticle to Killing spinors and to explore a limit in which the cosmological constant tends to zero and the electric and magnetic charges of the black hole vanish. This might give an answer to the question raised above. Second, in \cite{ag} an $\mathcal{N}=2$ superparticle was constructed in an {\it ad hoc} manner following the
recipe which was earlier applied to an $\mathcal{N}=4$ superparticle in
Bertotti--Robinson space \cite{bgik,gala2}. Below we give a proof that the $\mathcal{N}=2$ model is essentially unique and
is constructed in a purely group theoretical way.

The organization of this paper is as follows. In the next section
background fields are briefly discussed.
In Section 3 conformal mechanics associated with the near horizon extreme Kerr--Newman-AdS-dS black hole
is studied within the framework of the Hamiltonian formalism. First integrals corresponding to the $SO(2,1) \times U(1)$ isometry group are
identified. Section 4 is devoted to $\mathcal{N}=2$ supersymmetric extension of the conformal mechanics which is done within the canonical formalism.
Our analysis is fully general and shows that the $\mathcal{N}=2$  model is unique.
We summarize our results and discuss possible further developments in Section 5.

\vspace{0.5cm}

\noindent
{\bf 2. Background fields}\\

The Kerr-Newman--AdS--dS black hole is a solution of the Einstein-Maxwell equations with a non--vanishing cosmological constant \cite{carter}.
In Boyer--Lindquist--type coordinates it reads\footnote{We use metric with mostly minus signature and put $c=1$, $G=1$.
In these conventions the Einstein--Maxwell equations read
$R_{nm}-\frac 12 g_{nm} (R+2 \Lambda)=-2(F_{ns} {F_m}^s-\frac 14 g_{nm} F^2)$, $\partial_n (\sqrt{-g} F^{nm})=0$.}
\bea\label{kna}
&&
ds^2=\frac{\Delta_r}{\rho^2} \left(d t-\frac{a}{\Xi} \sin^2 \theta d \phi\right)^2
-\frac{\rho^2}{\Delta_r} d r^2 - \frac{\rho^2}{\Delta_\theta} d\theta^2
-\frac{\Delta_\theta}{\rho^2}\sin^2\theta\left(a d t-\frac{r^2+a^2}{\Xi} d \phi\right)^2,
\nonumber\\[2pt]
&&
A=-\frac{q_e r}{\rho^2}\left( dt - \frac{a\sin^2\t}{\Xi} d \phi \right)
-\frac{q_m\cos\t}{\rho^2}\left( a d t - \frac{r^2 + a^2}{\Xi} d\phi \right) ,
\eea
where
\bea
&&
\Delta_r=(r^2+a^2)\left(1+\frac{r^2}{l^2}\right)-2Mr+q^2, \qquad \Delta_\theta=1-\frac{a^2}{l^2}\cos^2\theta,
\nonumber\\[2pt]
&&
\rho^2=r^2+a^2\cos^2\theta,\qquad \Xi=1-\frac{a^2}{l^2},\qquad q^2=q_e^2+q_m^2.
\eea
The parameters $M$, $a$, $q_e$ and $q_m$ are linked to the mass, angular momentum, electric and magnetic charges of
the black hole, respectively (for explicit relations see e.g. \cite{strominger}). $l^2$ is taken to be positive for AdS and negative for dS and
is related to the cosmological constant via $\Lambda=-3/l^2$.

It is evident that (\ref{kna}) is invariant under the time translation and rotation around $z$--axis
\be\label{tp}
t'=t+\a, \qquad
\phi'=\phi+\b.
\ee
A less obvious fact is that this solution admits the second rank Killing tensor \cite{car,wp,carter1}
which obeys
\be\label{KT}
K_{mn}=K_{nm}, \qquad \nabla_{(n} K_{mp)}=0, \qquad {F_{n}}^m K_{mp}+{F_{p}}^m K_{mn}=0.
\ee
In Boyer--Lindquist--type coordinates it reads ($x^m=(t,r,\t,\phi)$)
\be\label{kt}
K_{mn}=Q_{mn}+r^2 g_{mn},
\ee
where
\be
Q_{mn}=\left(
\begin{array}{cccccc}
-\Delta_r & 0 & 0 & \frac{a \Delta_r }{\Xi} \sin^2{\t} \\
\nonumber\\[2pt]
0 & \frac{\rho^4}{\Delta_r} & 0 & 0\\
\nonumber\\[2pt]
0 & 0 & 0 & 0\\
\nonumber\\[2pt]
\frac{a \Delta_r}{\Xi}  \sin^2{\t} & 0 & 0 & -\frac{a^2 \Delta_r}{\Xi^2}  \sin^4{\t}\\
\end{array}
\right).
\ee
Generally,
a Killing tensor is a key ingredient in establishing complete integrability of the geodesic equations.
It also helps to separate variables for some important field equations
in a given gravitational background, including Klein--Gordon and Dirac equations (for a review and further references see e.g. \cite{fk}).
It should also be mentioned that, when considering superparticle models on a curved background which admits Killing--Yano tensors,
extra supersymmetry charges can be constructed \cite{grh} whose Poisson bracket yields Killing tensors\footnote{For some recent developments see  \cite{hh}.}.

The extreme solution is characterized by the condition that the inner and outer horizons of the black hole
coalesce. Denoting the corresponding value of the radial coordinate by $r_+$,
from $\Delta_r (r_+)=0$, $\Delta_r' (r_+)=0$ one finds
\begin{equation}
a^2 = \frac{r_+^2(1+3r_+^2/ l^2)-q^2}{1-r_+^2/ l^2},\qquad
M = \frac{r_+[(1+r_+^2/ l^2)^2-q^2/ l^2]}{1-r_+^2/ l^2},
\end{equation}
and
\be
\Delta_r(r)={(r-r_+)}^2[{(r+r_+)}^2+2 r_+^2+l^2+a^2]/l^2.
\ee

In order to describe the near horizon geometry, one redefines the coordinates \cite{strominger}
\be\label{red}
r \quad \rightarrow \quad r_+ + \epsilon r_0 r, \qquad t \quad \rightarrow \quad t r_0/\epsilon, \qquad
\phi \quad \rightarrow \quad \phi+
\frac{t r_0 a \Xi}{\epsilon (r_+^2+a^2)},
\ee
where $r_0$ is a constant to be fixed below
and then takes the limit $\epsilon \rightarrow 0$. The first prescription in (\ref{red}) followed by $\epsilon \rightarrow 0$
provides a natural definition of the near horizon region.
The last two prescriptions are designed so as to make $\lim_{\epsilon \rightarrow 0 } {d s}^2$ nonsingular\footnote{As to the vector potential, usually one
calculates the near horizon field strength and derives the vector potential from there.}
\bea\label{nhm}
&&
 ds^2=\Gamma(\theta)\left[
r^2dt^2-\frac{dr^2}{r^2}
-\alpha (\t) d\theta^2 \right] - \gamma(\theta)(d\phi+krdt)^2,
\quad A=f(\theta)(d\phi + k r dt),
\eea
where
\bea
&&
\Gamma(\theta)=\frac{\rho_+^2 r_0^2}{r_+^2+a^2}, \qquad \qquad
\alpha(\theta)=\frac{r_+^2+a^2}{\Delta_\theta r_0^2}, \qquad \qquad \quad
\gamma(\theta)=\frac{\Delta_\theta (r_+^2+a^2)^2\sin^2\theta}{\rho_+^2\Xi^2},
\nonumber
\eea
\bea
&&
\rho_+^2=r_+^2+a^2\cos^2\theta,\qquad
r_0^2=\frac{(r_+^2+a^2)(1-r_+^2/l^2)}{1+6r_+^2/l^2-3r_+^4/l^4-q^2/l^2}
,\qquad
k = \frac{2ar_+\Xi r_0^2}{(r_+^2+a^2)^2},\nonumber\\[2pt]
&&
f(\theta)= \frac{(r_+^2 + a^2)[q_e (r_+^2 - a^2\cos^2\t ) + 2q_m ar_+  \cos\t]}{2\rho_+^2 \Xi a r_+}.
\eea
This is a solution of the Einstein-Maxwell equations \cite{strominger} which generalizes
the near horizon extreme Kerr metric constructed in \cite{bh}. It reduces to the latter when $q_e=q_m=0$ and
$l^2\rightarrow\infty$.

Near the throat the symmetry group is enhanced.
In addition to (\ref{tp}) it includes the
dilatation
\be\label{tp1}
t'=t+\g t, \qquad r'=r-\g r,
\ee
and the special conformal transformation
\be\label{sp}
t'=t+(t^2+\frac{1}{r^2}) \s, \qquad r'=r-2 tr\s, \qquad \phi'=\phi-\frac{2k}{r} \s.
\ee
Altogether they form $SO(2,1) \times U(1)$ group \cite{strominger}.

Concluding this section we derive the second rank Killing tensor in the near horizon region.
Implementing the same limit as above one finds that
the second term in (\ref{kt}) reduces to the metric which gives a trivial contribution to the Killing tensor and
can be discarded. The first terms yields
\be\label{kt1}
K_{nm} d x^n d x^m=\Gamma(\t)^2 \left[r^2 dt^2-\frac{1}{r^2} dr^2 \right].
\ee
Given the background fields (\ref{nhm}), it is straightforward to verify that (\ref{kt1}) meets all the requirements formulated in (\ref{KT}).
Note that in a four--dimensional space--time with four isometries the Killing tensor can not be irreducible.
Reducibility of the Killing tensor is a priori guaranteed by the enhanced isometry. However, the explicit formula for the reduction of the Killing tensor into Killing vectors may be useful and is given below in Section 3. Note that up to a conformal factor (\ref{kt1}) coincides with the $AdS_2$ metric in Poincar\'e coordinates.

\vspace{0.5cm}

\noindent
{\bf 3. Conformal mechanics}\\

First it is worth briefly reminding a relation between symmetries of a background and conserved charges of a particle propagating in it.
The conventional action functional and equations of motion read
\be\label{AC}
S=-\int \left(m ds+e A \right) \quad \Rightarrow \quad m \left(\frac{d^2 x^n}{d s^2}+\Gamma^n_{mp} \frac{d x^m}{d s}\frac{d x^p}{d s}  \right)=e g^{nm} F_{mp}
\frac{d x^p}{d s},
\ee
where $m$ is the mass and $e$ is the electric charge of the particle.
In general, the coordinate transformation $x'^n=x^n+\xi^n(x)$ generated
by a Killing vector $\xi^n(x)$ leaves the background vector potential invariant if
\be\label{sc}
\xi^m F_{mn}+\partial_n (\xi^m A_m)=0
\ee
holds. Then from (\ref{AC}) and (\ref{sc}) one finds that
each Killing vector $\xi^n (x)$ gives rise to the integral of motion
\be\label{ekv}
\xi^n (x) (m g_{nm} \frac{d x^m}{d s}+e A_n),
\ee
which is linear in $\frac{d x^m}{d s}$. Likewise, each Killing tensor $K_{mn}$ yields the integral of motion
\be\label{cqk}
K_{nm} \frac{d x^n}{d s} \frac{d x^m}{d s},
\ee
which is quadratic in $\frac{d x^m}{d s}$. That (\ref{cqk}) is conserved is verified with the use of
the rightmost constraint in (\ref{KT}).

The action functional for a massive charged particle propagating near the horizon of the extreme Kerr-Newman--AdS--dS black
hole is
\be\label{start}
S=
-\int d t ~\left( m \sqrt{\Gamma(\t)\left[r^2-\dot r^2/r^2-\a(\t) ~\dot\t^2\right]-\g(\t) {\left[\dot\phi+k r \right]}^2}+e f(\t) \left[\dot\phi+k r \right]
 \right),
\ee
where the overdot denotes the derivative with respect to $t$.
We choose the Hamiltonian formalism to analyze the model. Introducing momenta $(p_r,p_\t,p_\phi)$ canonically
conjugate to the configuration space variables $(r,\t,\phi)$
\bea\label{momenta}
&&
p_r=\frac{m \Gamma(\t) \dot r}{r^2 \sqrt{\Gamma(\t)\left[r^2-\dot r^2/r^2-\a(\t) ~\dot\t^2\right]-\g(\t) {\left[\dot\phi+k r \right]}^2}},
\nonumber\\[2pt]
&&
p_\t=\frac{m \Gamma(\t) \a(\t) \dot \t}{\sqrt{\Gamma(\t)\left[r^2-\dot r^2/r^2-\a(\t) ~\dot\t^2\right]-\g(\t) {\left[\dot\phi+k r \right]}^2}},
\nonumber\\[2pt]
&&
p_\phi=\frac{m \gamma(\t) \left[\dot \phi+k r \right]}{\sqrt{\Gamma(\t)\left[r^2-\dot r^2/r^2-\a(\t) ~\dot\t^2\right]-\g(\t) {\left[\dot\phi+k r \right]}^2}}-e f(\t),
\eea
one readily finds the identity
\bea\label{iden}
&&
\sqrt{m^2 \Gamma(\t)+{(r p_r)}^2 +p_\t^2 / \a(\t)+\Gamma(\t) {[p_\phi+e f(\t)]}^2 / \g(\t)}=
\nonumber\\[2pt]
&& \qquad =
\frac{m \Gamma(\t) r}{\sqrt{\Gamma(\t)\left[r^2-\dot r^2/r^2-\a(\t) ~\dot\t^2\right]-\g(\t) {\left[\dot\phi+k r \right]}^2}}.
\eea
Then (\ref{momenta}) and (\ref{iden}) can be used to derive conserved charges, including the Hamiltonian $H$, the generator of special conformal transformations $K$, the generator of dilatations $D$
and the generator $P$ of rotations around $z$--axis, directly from the Killing vector fields (\ref{tp}), (\ref{tp1}), (\ref{sp}) and
the general formula (\ref{ekv})
\bea\label{hamc}
&&
H=r \left( \sqrt{m^2 \Gamma(\t)+{(r p_r)}^2 +p_\t^2 / \a(\t)+\Gamma(\t) {[p_\phi+e f(\t)]}^2 / \g(\t)} -k p_\phi \right),
\nonumber\\[2pt]
&&
K=\frac{1}{r} \left( \sqrt{m^2 \Gamma(\t)+{(r p_r)}^2 +p_\t^2 / \a(\t)+\Gamma(\t) {[p_\phi+e f(\t)]}^2 / \g(\t)} +k p_\phi \right)+
\nonumber\\[2pt]
&& \qquad +t^2 H+2trp_r, \qquad \qquad \quad  D=t H+r p_r, \qquad \qquad \quad P=p_\phi.
\eea
Under the Poisson bracket they form
$so(2,1)\oplus u(1)$ algebra
\be\label{confalg}
\{H,D \}=H, \quad \{H,K \}=2D, \quad \{D,K \} =K.
\ee

The Killing tensor (\ref{kt1})
yields the integral of motion quadratic in momenta
\be\label{L}
L=m^2 \Gamma(\t)+p_\t^2 / \a(\t)+\Gamma(\t) {[p_\phi+e f(\t)]}^2 / \g(\t).
\ee
This can be used to derive the explicit formula for the reduction of the Killing tensor into Killing vectors.
Computing the Casimir element of $so(2,1)$ one finds
\be\label{rela}
H K-D^2=L-k^2 P^2.
\ee
Translating this back to the Lagrangian formalism one can verify that
contributions linear in the vector potential
are canceled as well as those quadratic in the vector potential. The rest yields
\be
K_{nm}=\frac 12 \left(\xi^{(1)}_n \xi^{(3)}_m+\xi^{(3)}_n \xi^{(1)}_m \right)-\xi^{(2)}_n \xi^{(2)}_m+k^2 \xi^{(4)}_n \xi^{(4)}_m.
\ee
Here $(\xi^{(1)}_n,\xi^{(2)}_n,\xi^{(3)}_n,\xi^{(4)}_n)$ denote the Killing vectors corresponding to the time translation, dilatation, special conformal transformation and rotation around $z$--axis, respectively. As usual, the index is lowered with the use of the metric.
In view of the enhanced isometry similar formulae hold
for the extremal Kerr throat geometry \cite{ag}, for the near-horizon geometry of the extremal Kerr-NUT-AdS-dS
black hole \cite{jr}, and for the weakly charged extremal Kerr throat geometry \cite{fr}.

Finally, we note that Hamiltonian equations of motion can be treated by analogy with the extreme Kerr throat case \cite{ag}.
The dynamics of the radial pair $(r,p_r)$ is fixed from the conserved charges $H$ and $D$. The momentum $p_\phi$ is a constant of motion
while $p_\t$ is expressed algebraically via other variables with the use of $L$. The remaining equations of motion for $\t$ and $\phi$
can be integrated by quadratures only.

\vspace{0.5cm}

\noindent
{\bf 4. $\mathcal{N}=2$ superconformal mechanics}\\

In this section we construct $\mathcal{N}=2$ supersymmetric extension of the particle
propagating near the horizon of the extreme Kerr-Newman--AdS--dS black hole.
In addition to the $so(2,1)\oplus u(1)$ generators discussed in the previous section,
$su(1,1|1)$ superalgebra involves supersymmetry generators
$Q$, $\bar Q$, superconformal generators $S$, $\bar S$,
and $u(1)$ $R$--symmetry generator $J$. Here and in what follows the bar denotes complex conjugation.
The structure relations read\footnote{For a discussion of this algebra in the context of non--relativistic many--body mechanics see e.g. \cite{gl,ag1} and references therein.}
\begin{align}\label{sa}
&
\{Q,\bar Q \}=-2i H, && \{K,Q \}=S, && \{Q,\bar S \}=2i (D+i J),
\nonumber\\[2pt]
&
\{D,Q \}=-\frac 12 Q, && \{H,S \}=-Q, && \{D,S \}=\frac 12 S,
\nonumber\\[2pt]
&
\{S,\bar S \}=-2iK, && \{J,Q \}=-\frac i2 Q, && \{J,S \}=-\frac i2 S,
\nonumber\\[2pt]
&
\{H,D \}=H, && \{H,K \}=2D, && \{D,K \} =K,
\end{align}
plus complex conjugate relations.

A conventional way to accommodate $\mathcal{N}=2$ supersymmetry in the bosonic model under consideration is to introduce two fermonic degrees of freedom $\p$ and $\bar\p$
and impose the canonical bracket\footnote{When constructing a supersymmetric generalization, the action functional (\ref{start}) is extended by the fermionic kinetic term
$\int dt \Big(\frac{i}{2} \bar\p \dot \p-\frac{i}{2} \dot{\bar\p} \p \Big)$ as well as other contributions describing boson--fermion couplings. Within the framework of the Hamiltonian formalism there appear fermionic second class constraints,
(\ref{br}) being the Dirac bracket associated to them.}
\be\label{br}
\{\p,\bar\p \}=-i.
\ee
The most general form of the supersymmetry charges is
\be
Q=A e^{i B} \psi, \qquad \bar Q=A e^{-i B} \bar\p,
\ee
where $A$ and $B$ are real functions on the full phase space $(r,\t,\phi,p_r,p_\t,p_\phi)$ to be fixed from the structure relations of the superalgebra and the requirement that in the bosonic limit a supersymmetric Hamiltonian
reduces to the first line in (\ref{hamc}). From $\{Q,\bar Q \}=-2i H$ one finds
\be\label{fullH}
H=\frac 12 A^2+\frac 12 \{ A^2,B \} \p\bar\p,
\ee
while the bosonic limit gives
\be\label{A}
H_0=\frac 12 A^2=r \left( \sqrt{m^2 \Gamma(\t)+{(r p_r)}^2 +p_\t^2 / \a(\t)+\Gamma(\t) {[p_\phi+e f(\t)]}^2 / \g(\t)} -k p_\phi \right).
\ee
For a supersymmetric model the generators of dilatations and special conformal transformations can be chosen as in the preceding section
\be\label{DK}
D=t H+r p_r, \qquad K=K_0+t^2 H +2 t r p_r,
\ee
where we denoted
\be\label{K0}
K_0=\frac{1}{r} \left( \sqrt{m^2 \Gamma(\t)+{(r p_r)}^2 +p_\t^2 / \a(\t)+\Gamma(\t) {[p_\phi+e f(\t)]}^2 / \g(\t)} +k p_\phi \right),
\ee
and $H$ is now the full supersymmetric Hamiltonian (\ref{fullH}).

Given $Q$ and $K$, superconformal generator $S$ and $R$--symmetry generator $J$ are uniquely fixed by the
structure relations of the superalgebra (\ref{sa}). Thus it remains to compute all the brackets and derive restrictions on $B$.

The bracket of $D$ and $Q$ yields\footnote{Here and in what follows it proves helpful to use the brackets
$\{r p_r, H_0 \}=-H_0$, $\{r p_r, K_0 \}=K_0$, and $\{K_0,H_0 \}=-2 r p_r$ which reproduce the structure relations of $so(2,1)$.}
\be\label{r1}
\{r p_r,B \}=0,
\ee
which means that $B$ is a function of $(r p_r)$. Computing the bracket of $H$ and $K$ one gets
\be
\{K_0,\{H_0,B\} \}=0 \quad \Rightarrow \quad \{H_0,\{K_0,B \} \}=0,
\ee
which implies that $\{K_0,B \}$ is the integral of motion of the original bosonic theory.  In the last line the Jacobi identity was used.

The bracket of $K$ and $Q$ determines $S$
\be
S=\left(-t+i\{K_0,B\}-\frac{r p_r}{H_0} \right) Q,
\ee
while $\{S,\bar S \}=-2iK$ gives
\be\label{r2}
\{K_0,B \}^2=\frac{\left(H_0 K_0-{(r p_r)}^2 \right)}{H_0^2}=\frac{\left(L-k^2 p_\phi^2\right)}{H_0^2},
\ee
with $L$ from (\ref{L}). Note that the numerator coincides with the Casimir element of $so(2,1)$ realized in the bosonic theory.
The full fraction commutes with $H_0$ in agreement with what was found above.

Finally, computing the bracket of $Q$ and $\bar S$ one derives the $R$--symmetry generator
\be
J=H_0 \{ K_0,B\} +\left(1+\frac{K_0 \{H_0,B\}}{2 H_0 \{K_0,B \}} \right) \p\bar\p.
\ee
It turns out that the rest of the algebra does not impose further restrictions on the form of $B$.
Thus, in order to construct an $\mathcal{N}=2$ superparticle moving near the horizon of the extreme Kerr-Newman--AdS--dS black hole,
it is sufficient to solve (\ref{r1}) and (\ref{r2}) for $B$.

As was mentioned above, (\ref{r1}) means that $B$ depends on $r$ and $p_r$ in the form of the single argument $(r p_r)$. Taking into account the definition of $K_0$ from
(\ref{r2}) one derives a linear inhomogeneous first order partial differential equation for $B$. Its general solution is the sum of a particular solution to the inhomogeneous equation and the general solution to the homogeneous equation
\be\label{B}
B=-\arctan{\frac{r p_r}{\sqrt{L-k^2 p_\phi^2}}}+B_{hom}.
\ee
The latter is an arbitrary function of the integrals of the associated system of
ordinary differential equations (see e.g. \cite{smirnov}). It turns out that they all depend on $\phi$. Because rotation around $z$--axis is the symmetry of the bosonic model, the requirement that the momentum $p_\phi$ be conserved in the full supersymmetric theory rules out $B_{hom}$.

Thus, $\mathcal{N}=2$ supersymmetric extension is essentially unique.
Substituting (\ref{B}) in the generators above one gets
\begin{align}
&
H=H_0- \left( {K_0}^{-1} \sqrt{L-k^2 p_\phi^2}  \right) \p\bar\p, && J=\frac 12 \p\bar\p+\sqrt{L-k^2 p_\phi^2},
\nonumber\\[2pt]
&
D=t H+r p_r, && K=K_0+t^2 H +2 t r p_r,
\nonumber\\[2pt]
&
Q=-i {(K_0/2)}^{-1/2} \left(r p_r+i \sqrt{L-k^2 p_\phi^2} \right) \p, && S=-t Q+i {(2K_0)}^{1/2} \p.
\end{align}
Taking into account the fact that $H_0$, $K_0$, $r p_r$ obey the structure relations of $so(2,1)$ and
$H_0 K_0-{(r p_r)}^2=L-k^2 p_\phi^2$ is the Casimir element, one concludes that the $\mathcal{N}=2$ model is constructed in a purely group theoretical way.

A few comments are in order. First, the second term in the definition of $J$
might seem odd. It is the square root of the Casimir element of $so(2,1)$ which commutes with all the generators of
$su(1,1|1)$. If desirable, instead of including it in $J$, one could keep this term as the central charge in $\mathcal{N}=2$
superalgebra. Only the bracket of $Q$ and $\bar S$ would be altered. In fact, it is this way how the $R$--symmetry generator is usually
treated in the context of non--relativistic $\mathcal{N}=2$ many--body mechanics (see e.g. \cite{gl,ag1}). Second, by construction $p_\phi$ commutes with all the generators of $su(1,1|1)$. The full symmetry group of the superparticle is thus $SU(1,1|1) \times U(1)$.
Third, $\mathcal{N}=2$ supersymmetry does not impose any restriction on the particle parameters. This is to be contrasted with the
$\mathcal{N}=4$ superparticle moving near the horizon of the extreme Reissner--Nordstr\"om black hole where the higher symmetry leads to a kind of $BPS$ condition \cite{gala2}. Finally, our consideration implies that the $\mathcal{N}=2$ superparticle associated with the near horizon extreme Kerr geometry \cite{ag} is unique as well.

\vspace{0.5cm}

\noindent
{\bf 5. Conclusion}

\vspace{0.5cm}

To summarize, in this work we studied the conformal mechanics resulting from the near horizon geometry of the extreme
Kerr-Newman--AdS--dS black hole. Conserved charges of the particle were identified.
A unique $\mathcal{N}=2$ supersymmetric extension of the conformal mechanics was constructed
within the framework of the Hamiltonian formalism.

Proceeding to open questions, the first to mention is a link of the supersymmetry charges and the superconformal charges constructed in this work to Killing spinors characterizing the near horizon extreme Kerr-Newman--AdS--dS geometry. A related issue is a Lagrangian formulation
for the $\mathcal{N}=2$ superparticle both in components and in superfields. Then it is worth constructing
a canonical transformation to conventional conformal mechanics in the spirit of \cite{gala2}. The resulting angular sector might be interesting in the context of recent studies in \cite{hln} (see also references therein).
It is also intriguing to study a possible link of the model in this work to the matrix model for $AdS_2$ proposed in \cite{as}.
Finally, one could try to construct $\mathcal{N}>2$ supersymmetric extensions of the conformal mechanics, albeit it is not clear what they might correspond to in geometric terms.

\vspace{0.5cm}

\noindent{\bf Acknowledgements}\\

\noindent
This work was supported by the Dynasty Foundation and in part by grants RFBR 09-02-00078, LSS 3558.2010.2.

\end{document}